\documentclass[preprint,showpacs,preprintnumbers,amsmath,
    amssymb,showkeys,pre]{revtex4}

\usepackage{graphicx}
\usepackage{dcolumn}
\usepackage{bm}
\usepackage{amsthm}

\newcommand{\mn}{\mathbb{N}} 
\newcommand{\mz}{\mathbb{Z}} 
\newcommand{\var}[1]{\text{var}\left(#1\right)}
\newcommand{\Esym}{\text{E}}
\newcommand{\E}[1]{\Esym\left[#1\right]}
\newcommand{\Probsym}{\mathbb{P}}
\newcommand{\Prob}[1]{\Probsym\left[#1\right]}

\AtBeginDocument{\newenvironment{proctable}
{
                                                                                
\begin{quote} \begin{flushleft}\rule{8.5cm}{2pt}
 
}
{
 
\rule{8.5cm}{2pt}
 
\end{flushleft}
\end{quote} }}

\newtheorem{theorem}{Theorem}

\newtheorem{defn}{Definition}

\begin{document}
\title{A Markov Chain Based Method for Generating Long-Range Dependence}
\date{May 2005}

\author{Richard G Clegg}
\email{richard@richardclegg.org}
\author{Maurice Dodson}
\email{mmd1@york.ac.uk}
\affiliation{Department of Mathematics, University of York, York, UK. YO10 5DD}

\begin{abstract}
This paper describes a model for
generating time series which exhibit the statistical phenomenon known
as long-range dependence (LRD).  A Markov Modulated Process
based upon an infinite Markov chain is described.  The work described
is motivated by applications in telecommunications where LRD is a known 
property of time-series measured on the internet.  The process
can generate a time series exhibiting LRD with known parameters and
is particularly suitable for modelling internet traffic since the 
time series is in terms of ones and zeros which can be interpreted
as data packets and inter-packet gaps.
The method is extremely simple computationally and analytically
and could prove more tractable than other methods described
in the literature. 
\end{abstract}

\pacs{02.50.-r,89.75.Da, 05.10.-a,95.75.Pq,95.75.Wx}
\keywords{Long-Range Dependence, Scaling Phenomena }

\maketitle

\section{Introduction}

Long-range dependence (LRD) is a statistical phenomenon which is used 
to describe 
a process which exhibits significant correlations even between
widely separated points.  A more 
formal definition is given in the next part of this paper.  
Roughly speaking, a 
process with a high degree of LRD can be thought of as correlated at all
scales.  A good introduction to the topic of LRD is provided 
by \cite{beran1994} and a discussion in the context of telecommunications
traffic is given by \cite{clegg2004}.  LRD is most often characterised
by the Hurst parameter, $H$, which is in the range $(1/2,1)$ for a time
series which exhibits LRD.  If $H = 1/2$ then this indicates the
data is independent or has only short-range correlations.  The topic
of LRD has attracted a great deal of interest since LRD has been observed
in time series measured in fields as diverse as finance, internet traffic
and hydrology.  

This paper introduces and tests a mechanism for generating LRD based on
an infinite Markov chain.  The traffic stream generated is binary in 
nature and the model has only two parameters, the mean and the
Hurst parameter of the generated traffic.   
This section provides a brief introduction
to the topic
of LRD in the context of telecommunications networks and discusses currently
used generation mechanisms for modelling LRD and also methods which
are currently used to measure LRD in a time series.  Section
\ref{sec:model} describes the infinite Markov model.  In Section
\ref{sec:acf} it is proved that the model does in fact generate
a time series with a given mean and with LRD having a 
given Hurst parameter.  
Finally, Section \ref{sec:tests} tests the model against other standard
LRD generation models and discusses the advantages of the
model.

\subsection{A Brief Introduction to LRD}
 
A number of different (and not necessarily equivalent)
definitions of LRD are in use in the literature.  A commonly
used definition is the one given here.
\begin{defn}
A weakly stationary time series 
exhibits LRD if the absolute value of its autocorrelation 
function (ACF) $\rho(k)$ does not have a
finite sum.  That is,
\begin{equation*}
\sum_{k=-\infty}^{\infty} |\rho(k)| = \infty.
\end{equation*}
\label{defn:lrd}
\end{defn}
It is often assumed that the ACF has the specific asymptotic form,
\begin{equation}
\rho(k) \sim c_\rho k^{-\alpha},
\label{eqn:lrd}
\end{equation}
for some positive constant $c_\rho$ and some real $\alpha \in (0,1)$.
Note that this is equivalent to a functional form for the
spectral density $f(\lambda)$ defined by,
\begin{equation*}
f(\lambda)= \frac{\sigma^2}{2 \pi} \sum_{k= -\infty}^{\infty} 
\rho(k) e^{ik\lambda}.
\end{equation*}
Equation \eqref{defn:lrd} is equivalent to,
\begin{equation*}
f(\lambda) \sim c_f |\lambda|^{-\beta},
\end{equation*}
as $\lambda \rightarrow 0$,
where $\sigma^2$ is the variance, $c_f$ is some 
positive constant and $\beta \in (0,1)$.  The Hurst parameter is
then given by $H = (1 + \beta)/2$.

It should be noted that here, and throughout this paper, 
$f(x) \sim g(x)$ is used to mean $f(x)/g(x) \rightarrow 1$ as 
$x \rightarrow \infty$ --- sometimes, in the literature, the 
symbol is used to mean {\em asymptotically proportional to} or
$f(x)/g(x) \rightarrow k$ for some constant $k$ as $x \rightarrow \infty$.

The constant $\alpha$ in \eqref{eqn:lrd} is sometimes expressed in
terms of the Hurst parameter $H = 1 - \alpha/2$.  The Hurst parameter 
as defined by this relation and \eqref{eqn:lrd} is the most commonly
used measure of LRD in the telecommuncations literature.

The reason for the interest in the subject within the field
of telecommunications is the fact
that LRD has been observed in various time series related to
internet traffic \cite{leland1993, willinger1997, morris2000}.
It is widely recognised that the engineering implications of LRD on queuing
performance can be considerable.  If Internet traffic is not modelled
well by independent or short-range dependent models then much traditional
queuing theory work based upon the assumption of Poisson processes is
no longer appropriate.  Traffic which is long-range dependent in nature
can have a queuing performance which is significantly worse than
Poisson traffic.  Modelling has shown how phase transitions can arise
in computer networks \cite{ohira1998}, how  and this phase transition can
be related to LRD \cite{valverde2001,woolf2002}.

In general it has been found that a higher Hurst parameter
often increases delays in a network, increases
the probability of packet loss and
affects a number of measures of engineering importance.  In fact 
\citet{erramilli1996} claims that
the Hurst parameter is
``...a dominant characteristic for a number
of packet traffic engineering problems...''.  Some of the effects
on queuing performance are given by \cite{norros1994, sahinoglu1999}. 
However, \cite{neidhardt1998} shows
that while the Hurst parameter is important to queueing, the relationship
is not a simple one --- in some cases a high Hurst parameter may improve
performance or have no effect.  The issue of the scale and nature of the
effect of LRD on queuing remains contentious. 

\subsection{Current Generation Mechanisms for LRD}

A number of modelling techniques are currently used for generating
traffic streams exhibiting LRD. Of these, the most commonly encountered
in the telecommunications literature are Fractional Gaussian Noise processes
(FGN), Fractional  Auto-Regressive Integrated Moving Average models 
(FARIMA, also refered to as ARFIMA), iterated chaotic maps 
and wavelet modelling.  

The FGN process is usually defined as increments of the Fractional
Brownian Motion (FBM) process.  An FBM process 
$B_H(t)$ is defined by:
\begin{equation*}
\begin{split}
\Prob{B_H(t+k) - B_H(t) \leq x} = \\ (2 \pi) ^{-\frac{1}{2} }k ^{-H}
\int \limits_{-\infty}^{x} \exp\left(\frac{-u^2}{2k^{2H}}\right)du,
\end{split}
\end{equation*}
where $\Prob{X}$ is a probability of an event $X$ and
$H \in (1/2,1)$ is the Hurst parameter.  This can be seen as a
generalisation of the more common Gaussian White Noise process.  A number
of authors have described methods for generating FGN and FBM 
\cite{mandelbrot1971}, \cite{davies1987} and \cite{paxson1997}.

The FARIMA model is an obvious modification of the traditional ARIMA$(p,d,q)$
model from time series analysis, allowing
$d \in (-1/2, 1/2)$ instead of $d \in \mz_+$.  
FARIMA processes were proposed by \cite{granger1980} and a description in
the context of LRD can be found in \cite[pages 59--66]{beran1994}.
As might be expected the $d$ parameter relates
to the Hurst parameter.  The relation is simply  $H= d + 1/2$ --- note
that this only produces legitimate values for $H$ when $d \in (0,1/2)$. 

\begin{figure}
\begin{center}
\includegraphics[width=8.5cm]{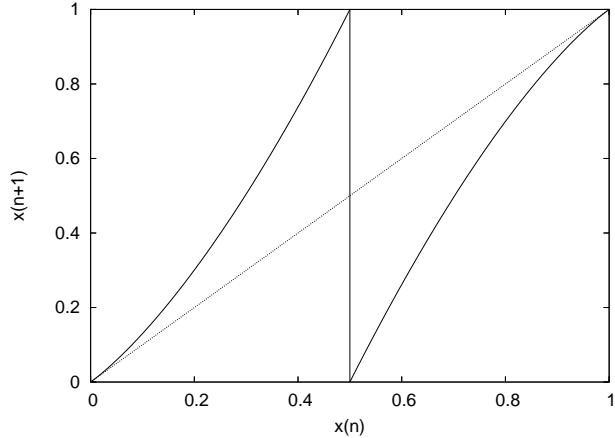}
\end{center}
\caption{Graph of a map which can be used to generate LRD.}
\label{fig:intmap}
\end{figure}

Iterated chaotic maps which exhibit intermittency
are also commonly used to generate time series
exhibiting LRD.  Given a starting value $x_0 \in (0,1)$
then a time series $\{x_n: n \in \mn\}$  can be generated
by the following map
\begin{equation*}
x_{n+1} =  
\begin{cases}
x_n+ \frac{1-d}{d^{m_1}} x_n^{m_1},  & 0 < x_n < d, \\
x_n- \frac{d}{(1-d)^{m_2}} (1 - x_n)^{m_2}, & d < x_n < 1,
\end{cases}
\label{eqn:chaoticmap}
\end{equation*}
where $d \in (0,1)$ and $m_1, m_2 \in (3/2,2)$.  If $x_0 \in (0,1)$
then $x_n \in (0,1)$ for all $n \in \mn$.  If this time series is
used to generate a binary time series $\{y_n: n \in \mn\}$ by the rule 
$y_i = 0$ if $x_i < d$ and $y_i = 1$ otherwise then the series
can be shown to exhibit LRD with a Hurst parameter given by
$H = (3m-4)/(2m-2)$.  This map is illustrated in Figure \ref{fig:intmap}.  An explanation 
for the presence of LRD in this map is provided by examining the behaviour
of the orbits at $x_i$ near zero or one.  The escape
from points near zero or one is extremely slow and this causes long
sequences of zeros or ones in the generated $y_i$ series.
Pioneering work in this area is \cite{wang1989} with early applications
to telecommunications being given by
\cite{erramilli1995}.  This mechanism is particularly suited for
generating data for modelling of packet networks since the ON ($y_i = 1$)
state can be considered to be a packet and the OFF ($y_i = 0$) state as
an interpacket gap.

In fact, the work described in \cite{wang1989} relates to the Markov chain
based work described in this paper as it approximates the chaotic map approach as
a piecewise linear map which can, in turn, be modelled as a Markov chain with
the topology described later.  A number of other papers have used 
Markov chains to model linear approximations to intermittent maps 
\cite{suetani1999,suetani2002,ashwin2004, giampieri2005}. 
The papers \cite{wang1989}, \cite{erramilli1995} 
and \cite{barenco2004} relate the piecewise linear approximations of intermittency 
maps to LRD and show how certain parameters for Markov chains give rise to
LRD in a process arising from the chain.  However, we can find no reference to papers 
which relate intermittency in general to LRD.

A technique gaining favour in modelling (and also in measuring) LRD is
wavelet analysis.  This allows the LRD hypothesis to be generalised
to multifractals.
LRD defines a single scaling behaviour for the
system (which applies in the tail of the ACF) --- 
if this scaling behaviour was the same at any scale then the process 
defined would be a monofractal.  However,
if the scaling behaviour differs across scales then the process is
multifractal.  There is some evidence that Internet traffic 
exhibits different
scaling behaviour at different timescales.  A general description of
multifractal processes and wavelets is found in \cite{riedi2003} and
a description of how wavelets can be used to create models with the
same multifractal spectrum as a given data set can be found in 
\cite{riedi1999}.

\subsection{Measurement Techniques for LRD}

A number of techniques are known for estimating the Hurst parameter from
real data.  There is no single technique which can be considered
perfect.  Computer code and analysis of various techniques can be
found at \cite{taqquwww}.  Comparisons of measurement techniques can
be found in \cite{taqqu1995}, \cite{taqqu1997} and \cite{bardet2003}.
In this paper, five techniques are used: the R/S statistic (in two
variants), the Aggregated Variance, the Periodogram, Whittle's Local
Estimator and a wavelet based technique.

The R/S statistic (also known as rescaled adjusted range) is one of 
the oldest
and best known techniques for estimating $H$.  It is a time domain
method which relies on considering the way that $R/S(n)$ varies 
with $n$ where $R$ is
the range, $S$ is the sample variance and $n$ is a scale (sample size)
within the time series. 
It is discussed in
detail in \cite{mandelbrot1969} and also \cite[pages 83--87]{beran1994}. 
There are several problems with this technique which are cited in the
literature.  The estimate produced is highly sensitive to the range
of scales examined.   In this
paper
two versions of the estimator are used which choose the 
scales to investigate in different ways.
The estimator is known to be biased and also slow
to converge.  It is included in this paper mainly for its historic
importance since it has become a standard measure despite its known
weaknesses.

The Aggregated Variance estimator produces an estimate for the Hurst
parameter by considering how the variance of the time series scales
as the series itself is aggregated into blocks.  Again this is a
time domain technique with known weaknesses --- jumps in the mean
and slowly decaying trends in particular can be issues.  
A fuller description can be found in \cite[page 92]{beran1994}.

The periodogram is one of the oldest frequency domain based estimators
and is described in \cite{geweke1983}.  It involves producing an estimate
for the spectral density $I(\lambda)$ of the time series and considering 
the slope of this as $|\lambda| \rightarrow 0$.  Theoretically, for
LRD, a log-log plot of the periodogram should have a slope of $1 - 2H$ close
to the origin.

Whittle's estimator \cite{fox1986} is a frequency domain 
technique which uses an approximate maximum likelihood estimator and 
an estimated functional form for the spectral density $I(\lambda)$
based upon an assumed underlying model.  Here, the Local Whittle variant is
used \cite{robinson1995} which is a semi-parametric version assuming
a functional form for $I(\lambda)$ only as $|\lambda| \rightarrow 0$.

Wavelet analysis has already been mentioned as a modelling technique
and has been used for the estimation of the Hurst parameter. 
In addition this has the benefit of providing an estimate of the
multifractal spectrum of the data  \cite{riedi2003, riedi1999}.
This method is based upon considering the behaviour of the frequency
spectrum although wavelets themselves are a technique to allow insight
into both frequency and time-domain behaviour simultaneously.

\subsection{The Need for a Parsimonious and Tractable LRD 
Generation Method}

Given the large (and not exhaustive) list of modelling techniques
already mentioned, it might be asked whether there
is a need for another model.  However, the model here is specifically
designed to be the simplest possible computational model which produces
LRD.

Fractional Gaussian Noise and
FARIMA are relatively simple to analyse from 
a statistical point of view (though the model described here is arguably
simpler).  However, these processes 
cannot easily be calculated in an ongoing manner (that is, the entire 
time series is usually generated ``at once'' and, having generated
$n$ points, the user must effectively start again to generate the $n+1$th 
point).

Iterated chaotic maps are computationally
parsimonious but are analytically problematic since no closed form
for the invariant density of the map described in the previous section
is known.  Therefore, it is difficult
to generate traffic with a given mean using the iterated map method and 
progress theoretically is difficult. An intermittent map with a known 
invariant density is given by \cite{artuso2003}, however, it is not known 
if this map would generate LRD and other barriers exist to computational
implementation.

The generation mechanism given here is extremely simple, theoretically
sound and has only two parameters, the mean and the Hurst parameter.
The data produced is produced in a stream of ones and zeros
and can be simply used with simulation models of networks --- the one
representing a data packet and the zero representing an inter-packet gap.

\section{The Markov Model for LRD}
\label{sec:model}

\begin{figure}[htb]
\begin{center}
\includegraphics[width=8.5cm]{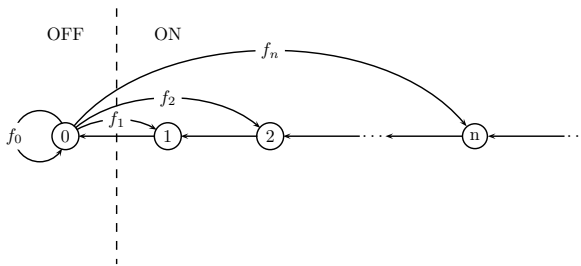}
\end{center}
\caption{An infinite Markov chain which 
generates a time series exhibiting LRD.}
\label{fig:markov_model}
\end{figure}

Figure \ref{fig:markov_model} shows an infinite Markov chain which
can be used to generate a time series exhibiting LRD.  This
particular chain with different transition probabilities
has been studied by a number of authors, notably, in this
context \citet{wang1989} and \citet{barenco2004} (the latter
also investigates the
double sided version).  The parameters $f_i$ are the transition
probabilities for reaching a given state $i$ from state $0$.  Also
$\pi_i$ is defined as the equilibrium probability of state $i$.  It
is clear that $\sum_{i=0}^\infty f_i = 1$ and also that 
$\sum_{i=0}^\infty \pi_i = 1$.  
More details and expanded versions
of the proofs included here can be found in \cite[Chapter 2]{clegg2004}.

The chain shown,  given a starting state $X_0 \in \mz_+$, produces a
Markov time series $\{X_i : i \in \mn\}$ where all the $X_i \in \mz_+$.  In
turn, this chain can generate
a time series $\{Y_i : i \in \mn\}$ where $Y_i = 0$ if
$X_i = 0$ and $Y_i = 1$ otherwise.  

It can be easily shown that the chain above is ergodic (and hence
the equilibrium distribution exists) if $\sum_{i=0}^\infty i f_i < \infty$
and also $\forall i \in \mn,\exists j > i : f_j > 0$ --- 
the first condition
ensures that the mean return time to the zero state is finite, the
second ensures that any state in the chain can be reached from the zero
state (obviously the zero state will be reached from any state $i$ in
exactly $i$ steps).  For the rest of this paper it will be assumed that
any chain discussed meets these conditions for ergodicity.

\begin{theorem}
The equilibrium distribution of the $i$th state is given by,
\begin{equation*}
\pi_i =  \pi_0 \sum_{j=i}^\infty f_j.
\end{equation*}
\end{theorem}
\begin{proof}
For a state $i$ then at equilibrium, the inputs to a state will 
sum to $\pi_i$.  That is,
\begin{equation}
\pi_i= \pi_{i+1} + \pi_0 f_i.
\label{eqn:pii}
\end{equation}
Substituting the same equation for $\pi_{i+1}$ gives,
\begin{equation*}
\pi_i= \pi_0 f_i + \pi_0 f_{i+1} + \pi_{i+2},
\end{equation*}
and repeating this subsitution recursively gives the proof.
\end{proof}
Note that since $\sum_{j=0}^\infty f_j= 1$ then for $i=0$ this equation
simply says $\pi_0 = \pi_0$.  Since all the $\pi_i$ must sum to one then,
in addition,
\begin{equation*}
\pi_0 = 1 - \sum_{i=1}^\infty \pi_i = (1 + \sum_{i=1}^ \infty i f_i)^{-1},
\end{equation*}
which, as has already been discussed, is finite.

\subsection{Introducing LRD into the model}

LRD with Hurst parameter $H$ can be guaranteed if the ACF $\rho(k)$
meets the condition given by \eqref{eqn:lrd}.
The most obvious way to induce a correlation for a lag $k$
into such a model is to choose the
$f_i$ in such a way that unbroken sequences of $k$ or more ones 
occur in the $Y_n$ series with the required frequency.  
Therefore, it would be suspected that the condition,
\begin{equation*}
\Prob{Y_i=1, Y_{i+1} = 1 \dots Y_{i+k} = 1} \sim C k ^ {-\alpha},
\end{equation*}
where $\alpha \in (0,1)$
will produce LRD with $H= 1 -\alpha / 2$.  To meet this requirement,
the following strict condition is introduced for $k > 0$,
\begin{equation*}
\sum_{i=k}^\infty \pi_i = C k ^ {-\alpha},
\end{equation*}
where $C$ is a constant.  Note that there is no guarantee that
this is a valid Markov chain --- conditions for this will be given later.  
By  setting $k=1$ it is immediate that $C = 1 - \pi_0$.  This gives,
\begin{equation*}
\sum_{i=k}^\infty \pi_i = (1 - \pi_0) k ^ {-\alpha} \qquad k > 0.
\end{equation*}
Subtracting the equation for $k+1$ gives,
\begin{equation*}
\pi_k = (1 - \pi_0) [ k^ {-\alpha} - (k+1) ^ {-\alpha}] \qquad k > 0.
\end{equation*}
From \eqref{eqn:pii} for $k$ then,
\begin{equation*}
\pi_0 f_k = \pi_k - \pi_{k+1} \qquad k > 0,
\end{equation*}
and therefore for $k > 0$,
\begin{equation}
f_k = \frac{1 - \pi_0} {\pi_0} \left[ k ^ {-\alpha} 
    - 2 (k+1) ^ {-\alpha}
    +(k+2) ^ {-\alpha}\right],
\label{eqn:f_k}
\end{equation}
and also, since $f_0 = 1 - \sum_{i=1}^\infty f_i$,
\begin{equation*}
f_0= 1 - \frac{1 - \pi_0} {\pi_0} \left[\sum_{i=1}^{\infty} 
     i ^ {-\alpha} - 2 \sum_{i=2}^{\infty} i ^ {-\alpha}
    + \sum_{i=3}^{\infty}i ^ {-\alpha} \right].
\end{equation*}
Most of the terms of the sum cancel leaving
\begin{equation}
f_0= 1 - \frac{1 - \pi_0} {\pi_0} \left[ 1 - 2^{-\alpha}\right].
\label{eqn:f0_final}
\end{equation}

The two equations, \eqref{eqn:f_k} and \eqref{eqn:f0_final} form the
model for LRD.  The model is defined by two parameters $\pi_0$ and
$\alpha$.  The $\alpha$ parameter is related to the Hurst parameter
as shown.  The $\pi_0$ parameter is the equilibrium probability of
state zero.  Hence $1 - \pi_0$ is the sum of all other equilibrium
probabilities and, therefore, the probability that any given $Y_i = 1$.
Therefore, the expectation value of $Y_i$ is given by, $\E{Y_i} = 1 - \pi_0$.
It remains to be proved that the model does generate LRD with the
required Hurst parameter and this is shown in the next section of
the paper.

It can be easily shown that this model meets the conditions for
ergodicity established earlier.  However, 
it should be noted that this model is not valid for every possible
combination of $\pi_0$ and $\alpha$.  In particular, for values of $\pi_0$ 
near zero then the term $(1 - \pi_0)/\pi_0$ becomes large
and values of $f_i$ from \eqref{eqn:f_k} will be negative, a
contradiction since the $f_i$ are probabilities.  
The fact that the model is invalid for
some combinations of $\pi_0$ and $\alpha$ simply means that for practical
experiments the
model must be confined to the valid region.  Rearranging
equation \eqref{eqn:f0_final} shows that for $\alpha, \pi_0 \in (0,1)$ then
$f_0 \in (0,1)$ if
\begin{equation*}
\pi_0 > \frac{2^\alpha - 1}{2^{\alpha+1} -1},
\end{equation*}
and this defines a valid region for the model.

\section{The ACF of the Markov Model}
\label{sec:acf}

It must now be shown that the model described in the previous section
does produce traffic with a given mean and Hurst parameter.  To recap,
the model relies on a Markov chain of the form shown in Figure
\ref{fig:markov_model} and with transition probabilities given by
\eqref{eqn:f_k} and \eqref{eqn:f0_final}.  The parameters of the
model are $\pi_0$ and $\alpha$.  Given some starting $X_0 \in \mz_+$, 
the Markov chain produces a time
series $\{ X_i: i \in \mn\}$ where $X_i$ is the state of the chain
at the $i$th iteration.  This is used to produce another time series
$\{ Y_i: i \in \mn\}$ where $Y_i = 0$ if $X_i = 0$ and $Y_i = 1$ otherwise.
(The time series $Y_i = 1$ if $X_i = 0$ and $Y_i = 0$ otherwise
also produces a series with LRD and mean $\pi_0$.)
This series has LRD with mean $1 - \pi_0$ and Hurst parameter 
$H = 1 - \alpha/2$.  That $\E{Y_i} = 1 - \pi_0$ has already been
shown.  It remains to show that the series has an ACF which follows
the form in \eqref{eqn:lrd} and this requires a result due to
\citet{feller1949} and is based on \citet{wang1989}.

\subsection{Proof that the Chain Generates LRD}

The event $\varepsilon$ occurs whenever $X_i = 0$.
It can easily be seen that the number of samples between successive
occurrences of $\varepsilon$ is an independent and identically
distributed variable and hence meets the definition in \cite{feller1949}.
A ``trial'' in the terms of \cite{feller1949} is equivalent to one
iteration of the Markov chain in this model.

\begin{defn}
If $\varepsilon$ occurs at the zeroth trial then let
the number of occurrences of $\varepsilon$ in
$n$ trials be $N_n$.
Let $F(n)$ be the distribution function of the number of trials between 
one event $\varepsilon$ and the next.  (Note that these definitions 
are those used by \cite{feller1949}).
\end{defn}

 If the event $\varepsilon$ has just
occurred (at the zeroth trial)
then the chain is in state 0.  If the chain makes the transition 
to state $i-1$ then
the event $\varepsilon$ will occur in $i$ steps.  Therefore, 
the distribution function is
given by
\begin{equation}
F(n)= \sum_{i=1}^n f_{i-1}.
\label{eqn:chainpdf}
\end{equation}

The results from \cite{feller1949} and \cite{wang1989} both assume
that the distribution function $F(n)$ obeys
\begin{equation}
1 - F(n) \sim A n ^ \gamma,
\label{eqn:fellercond}
\end{equation}
for some positive constant $A$ and some $\gamma$.  This will now
be shown for the specified infinite chain.

From equation \eqref{eqn:chainpdf},
\begin{equation*}
1-F(n)= 1 - \sum_{i=1}^n f_{i-1} = \sum_{i=n+1}^\infty f_{i-1} =
\sum_{i=n}^\infty f_i.
\end{equation*}
Substituting $f_i$ from \eqref{eqn:f_k}:
\begin{align*}
1-F(n) & = \left( \frac{1- \pi_0}{\pi_0} \right) \\
& \quad \sum_{i=n}^\infty 
\left[ i ^ {-\alpha} - 2 (i+1) ^{-\alpha} + (i+2) ^{-\alpha}
\right] \\
 & = \left(\frac{1- \pi_0}{\pi_0}\right) \left [
n ^ {-\alpha} - (n+1) ^ {-\alpha} \right] \\
& = \left(\frac{1- \pi_0}{\pi_0}\right) \frac { (n+1) ^ \alpha - n ^ \alpha }
{(n+1)^ \alpha n ^ \alpha} \\
& =  \left(\frac{1- \pi_0}{\pi_0}\right) \frac { (1 + 1/n) ^ \alpha - 1}
{n^\alpha(1 + 1/n) ^ \alpha}.
\end{align*}
Expanding $(1 + 1/n)^ \alpha$ using the binomial theorem gives
\begin{equation*}
(1+1/n) ^ \alpha = 1 + \alpha/n + O(n^{-2}).
\end{equation*}
Substituting this expression top and bottom gives
\begin{align*}
1-F(n) & = \left(\frac{1- \pi_0}{\pi_0}\right) 
\frac{ 1 + \alpha/n  + O(n^{-2}) - 1}
{ n^ \alpha(1 + \alpha/n + O(n^{-2}))} \\
& = \left(\frac{1- \pi_0}{\pi_0}\right) \frac{ n^{-\alpha}(\alpha/n + O(n^{-2}))}
{(1 + \alpha/n + O(n^{-2}))} \\
& \sim \left(\frac{1- \pi_0}{\pi_0}\right) \alpha n^{-(1+\alpha)},
\end{align*}
where  $f(n) = O(g(n))$ for functions $f(n)$ and $g(n)$ means that 
$|f(n)| < A g(n)$ for some positive constant $A$ and all $n > 0$.
This is the form required by equation \eqref{eqn:fellercond}
with $\gamma = (1 + \alpha)$ and $A= \alpha(1 - \pi_0) / \pi_0$.

From \cite[Theorem 10]{feller1949}, given that the probability 
distribution satisifies $1 - F(x) \sim Ax^{-\gamma}$, where $A$
is a positive constant and 
$1 < \gamma < 2$ then
\begin{equation*}
\var{N_n} \sim \frac{2A}{(2 - \gamma)(3 - \gamma) \mu^3} n^{3 - \gamma}.
\end{equation*}
In the case of the chain under investigation $\gamma = 1 + \alpha$,
and $A= \alpha(1- \pi_0)/\pi_0$.  Since the chain is ergodic,
the mean recurrence time of state zero for the infinite chain is
$1/\pi_0$.  Therefore,
\begin{equation}
\var{N_n} \sim \frac{2 \alpha \pi_0^2(1 - \pi_0)}
{(1 - \alpha)(2 - \alpha)}n^{2 - \alpha}.
\label{eqn:varnn}
\end{equation}

From \cite{wang1989} (equations 2.26a and 2.26c) if 
\begin{equation*}
\var{N_n}\sim K n^{2 - \alpha},
\end{equation*}
for some positive constant $K$ and some $\alpha \in (0,1)$ then the
autocorrelation function is given by
\begin{equation*}
\rho(n)\sim C n ^ {-\alpha},
\end{equation*}
where $C$ is some positive constant.  This is the form required by
\eqref{eqn:lrd}.

\section{Tests on the Markov Model}
\label{sec:tests}

In this section, two standard models for generating LRD are
compared with the Markov model described in this paper.  The computational
performance of the algorithm is compared against other algorithms.

\subsection{Practical Implementation of the Model}

The only difficulty in modelling the situation on a computer
comes in calculating $X_{n+1}$ 
when $X_n= 0$.  In this case, a random number generator and
the transition probabilities $f_j$ must be used to find the next 
state.  A naive approach to this would be to generate a random
number $r$, uniformly distributed in $(0,1)$ and say that $X_{n+1}$ is the
smallest $i$ such that $\sum_{j=0}^i f_i < r$.  This is fine for
low values of $i$ but as $i$ increases then this procedure becomes
inaccurate due to the finite precision arthmetic used by
computers.  The problem is that, as $i$ increases the
sum gets nearer to one but the $f_i$ get nearer to zero (since
adding numbers approaching zero to numbers approaching one is likely
to produce severe rounding error problems).  Hence
the errors in each stage of addition get larger.  However,
by the very nature of LRD, large values of $i$ are very likely
to come up.

It can simply be shown that
if $X_n = 0$ and  $0 < k \leq i \leq j$,
\begin{equation}
\begin{split}
\Prob{X_{n+1} \in [i,j] | X_{n+1} \in [k, \infty]}  =  \\
\frac{i ^ {-\alpha} - (i+1) ^ {-\alpha}  - 
(j+1) ^ {-\alpha} +(j+2) ^ {-\alpha}} 
{k ^ {-\alpha} - (k+1) ^ {-\alpha}}.
\label{eqn:PrXinfrange}
\end{split}
\end{equation}
Using this equation, Table \ref{tab:infchain} shows a procedure
for generating the sequence $\{X_n: n \in \mn\}$ given some
randomly chosen $X_0$.

\begin{table}[htbp!]
\begin{proctable}
\begin{enumerate}
\item If $X_n > 0$ then $X_{n+1} = X_n -1$.  Exit here.
\item Explicitly calculate $\Prob{X_{n+1} > j}$  
for values of $j \leq N$ where $N$ is some
small integer.  Use the procedure for 
the finite state model to find a value for
$X_{n+1}$ if $X_{n+1} < N$.
\item Generate a new random number $R$ 
in the range $[0,1]$.
\item Calculate 
$\Prob{X_{n+1} \in [N,2N-1] | X_{n+1} \in [N, \infty]}$ 
from equation \eqref{eqn:PrXinfrange}.
If $R$ is less than or equal to this probability 
then $X_{n+1}$ is in the required range.  Otherwise
go to step six.
\item If $X_{n+1}$ is in the required range then 
refine down by generating a new random
number and seeing if $X_{n+1}$ is in the range 
$[N, (3/2)N]$.  Continue refining by a
binary search (with a new random number 
each time) until $X_{n+1}$ is found.  Exit here.
\item Increase the value of $N$ to $2N$ and go to step 3.
\end{enumerate}
\end{proctable}
\caption{A procedure for finding $X_{n+1}$ from $X_n$ in
the infinite chain.}
\label{tab:infchain}
\end{table}

\subsection{Hurst Parameter Estimates}

Three generation mechanisms for LRD are compared, Fractional
Gaussian Noise (FGN), iterated maps (it. map) and the Markov
method developed in this paper.  For each method, three
different Hurst parameters are investigated and for each of
these, three data realisations are created.  For each 
realisation, one million points were generated (in the case
of the iterated map and Markov method, each of those points
was an aggregate of one hundred zeros and ones).
The Hurst
parameter was  estimated using the previously
discussed measurement techniques to check the match between theory 
and experiment.

The three methods
were implemented in the C programming language.  On a 2GHz processor
PC running Debian linux, to generate one 
million points  took 55 seconds for the Markov method, 
60 seconds 
for the iterated
maps method and 6 seconds for the fractional Gaussian noise method.  
However, it is debatable whether this is a fair comparison since the 
first two methods could be considered to be generating a hundred 
million points and aggregating into groups
of one hundred.  No C code to generate FARIMA based data was 
available and the R code 
available took 188 seconds to generate only a hundred thousand points ---
the run time did not seem to scale linearly and the test to generate
a million points was stopped after several hours.

\begin{table}
\begin{center}
\begin{tabular}{|l l|l l l | l l l | } \hline

Source & H & R/S & Mod. &  Agg.  & Period- & 
Local  & Wave- \\ 
 &  &  &  R/S &  Var. & ogram & 
 Whit. & lets \\ \hline
FGN & 0.625 & 0.637 & 0.624 & 0.623 & 0.626 & 0.639 & 0.635 \\
FGN & 0.625 & 0.632 & 0.624 & 0.622 & 0.624 & 0.638 & 0.635 \\
FGN & 0.625 & 0.645 & 0.633 & 0.620 & 0.622 & 0.638 & 0.635 \\ \hline
FGN & 0.75 &  0.728 & 0.738 & 0.741 & 0.747 & 0.774 & 0.767 \\
FGN & 0.75 &  0.741 & 0.736 & 0.749 & 0.755 & 0.776 & 0.769 \\
FGN & 0.75 &  0.694 & 0.719 & 0.741 & 0.754 & 0.774 & 0.768 \\ \hline
FGN & 0.875 & 0.784 & 0.837 & 0.858 & 0.877 & 0.908 & 0.897 \\
FGN & 0.875 & 0.750 & 0.823 & 0.850 & 0.876 & 0.908 & 0.897 \\
FGN & 0.875 & 0.747 & 0.835 & 0.860 & 0.876 & 0.908 & 0.898 \\ \hline
It. map & 0.625 & 0.635 & 0.590 & 0.604 & 0.630 & 0.719 & 0.706 \\
It. map & 0.625 & 0.608 & 0.595 & 0.604 & 0.627 & 0.716 & 0.703 \\
It. map & 0.625 & 0.637 & 0.594 & 0.610 & 0.637 & 0.718 & 0.707 \\ \hline
It. map & 0.75  & 0.828 & 0.666 & 0.717 & 0.746 & 0.813 & 0.800 \\
It. map & 0.75  & 0.725 & 0.650 & 0.712 & 0.739 & 0.813 & 0.801 \\
It. map & 0.75  & 0.678 & 0.694 & 0.765 & 0.768 & 0.814 & 0.803 \\ \hline
It. map & 0.875 & 0.703 & 0.779 & 0.851 & 0.876 & 0.925 & 0.910 \\
It. map & 0.875 & 0.779 & 0.802 & 0.854 & 0.877 & 0.924 & 0.910 \\
It. map & 0.875 & 0.846 & 0.817 & 0.861 & 0.874 & 0.925 & 0.912 \\ \hline
Markov & 0.625 & 0.526 & 0.597 & 0.611 & 0.621 & 0.703 & 0.691 \\
Markov & 0.625 & 0.593 & 0.645 & 0.700 & 0.684 & 0.710 & 0.702 \\
Markov & 0.625 & 0.632 & 0.603 & 0.646 & 0.650 & 0.707 & 0.698 \\ \hline
Markov & 0.75  & 0.663 & 0.684 & 0.744 & 0.760 & 0.793 & 0.784 \\
Markov & 0.75  & 0.670 & 0.667 & 0.751 & 0.759 & 0.793 & 0.783 \\
Markov & 0.75  & 0.671 & 0.671 & 0.724 & 0.736 & 0.786 & 0.776 \\ \hline
Markov & 0.875 & 0.724 & 0.732 & 0.816 & 0.848 & 0.884 & 0.873 \\
Markov & 0.875 & 0.757 & 0.754 & 0.830 & 0.859 & 0.885 & 0.874 \\
Markov & 0.875 & 0.656 &  0.781 & 0.852 & 0.866 & 0.885 & 0.875 \\ \hline
\end{tabular}
\end{center}
\caption{Hurst Parameter Estimates on Simulated Data.}
\label{tab:hurstests}
\end{table}

It would naturally be expected that the FGN model is the easiest to estimate
and this shows in the results in Table \ref{tab:hurstests}.  All the estimators
were relatively close to correct with the possible exception of the R/S plot 
on traffic with a Hurst parameter of 0.875 where the underestimate of $H$ was
quite severe.

Estimates on the iterated chaotic map traffic were not so successful.  The raw
R/S plot proved inconsistent and had a hard time estimating higher hurst parameters.  
It should be noted, for example, that for $H = 0.75$ estimates varied from 0.678 
to 0.828.  The performance for $H = 0.875$ was similarly bad.  The modified R/S
parameter was better in that it was more stable across runs but tended to
overestimate.  Local Whittle and wavelets tended to overestimate the Hurst parameter.
It should also be noted that the true result was regularly outside the  
95\% confidence intervals
for the wavelet estimator.

Estimates for the Markov based method were, in many ways, similar to the iterated
map method.  If anything, the results from the estimators are slightly closer to the
theory and this is particularly notable for the wavelet and local Whittle case.  
The evidence provided by the estimators is hard to interpret.  However, it can
certainly be said that the results for the Markov method are as close as the
results for the iterated map method.

Generally, considering the estimators themselves, the R/S method seemed unreliable 
(and this agrees with theory which shows it to be a biased estimator with poor
convergence).  The local Whittle and wavelets methods which have better theoretical
backing seem to have a better agreement with theory but it is worrying that the true
Hurst parameter for the data lay outside 95\% confidence for the wavelet estimator in
many cases.

\section{Conclusions}

The method for generating LRD shown here is computationally efficient,
extremely simple and produces a data stream with a given mean and Hurst
paramter.  The data stream can be generated in an online
manner (that is, the method can be started without knowing how many
points must ultimately be generated unlike, for example, FGN).  
The method has been proved theoretically to generate LRD with the
required parameters and this has been tested against a variety of
known estimators for the Hurst parameter.  It is interesting
to see quite how badly certain estimators perform even against 
very standard LRD generation mechanisms.

Compared with existing methods of generating LRD this procedure has
a number of extremely attractive properties.  It is computationally
and mathematically extremely simple.  While other models may
have more flexibility for precisely representing the nature of
the time series being simulated, it is hard to imagine
a simpler model for generating LRD.  It is hoped, therefore, that
this model will be tractable analytically for further developments,
for example, analysis of queuing performance of traffic generated
by such a model.

\begin{acknowledgments}
The authors would like to thank David Arrowsmith and 
Martino Barenco for their contributions to this research.
\end{acknowledgments}

\bibliography{rgcpre2004}

\begin{thebibliography}{35}
\expandafter\ifx\csname natexlab\endcsname\relax\def\natexlab#1{#1}\fi
\expandafter\ifx\csname bibnamefont\endcsname\relax
  \def\bibnamefont#1{#1}\fi
\expandafter\ifx\csname bibfnamefont\endcsname\relax
  \def\bibfnamefont#1{#1}\fi
\expandafter\ifx\csname citenamefont\endcsname\relax
  \def\citenamefont#1{#1}\fi
\expandafter\ifx\csname url\endcsname\relax
  \def\url#1{\texttt{#1}}\fi
\expandafter\ifx\csname urlprefix\endcsname\relax\def\urlprefix{URL }\fi
\providecommand{\bibinfo}[2]{#2}
\providecommand{\eprint}[2][]{\url{#2}}

\bibitem[{\citenamefont{Beran}(1994)}]{beran1994}
\bibinfo{author}{\bibfnamefont{J.}~\bibnamefont{Beran}},
  \emph{\bibinfo{title}{Statistics For Long-Memory Processes}}
  (\bibinfo{publisher}{Chapman and Hall}, \bibinfo{year}{1994}).

\bibitem[{\citenamefont{Clegg}(2004)}]{clegg2004}
\bibinfo{author}{\bibfnamefont{R.~G.} \bibnamefont{Clegg}}, Ph.D. thesis,
  \bibinfo{school}{Dept. of Math., Uni. of York., York.}
  (\bibinfo{year}{2004}), \bibinfo{note}{available online at: \\ {\tt
  www.richardclegg.org/pubs/thesis.pdf}}.

\bibitem[{\citenamefont{Leland et~al.}(1993)\citenamefont{Leland, Taqqu,
  Willinger, and Wilson}}]{leland1993}
\bibinfo{author}{\bibfnamefont{W.~E.} \bibnamefont{Leland}},
  \bibinfo{author}{\bibfnamefont{M.~S.} \bibnamefont{Taqqu}},
  \bibinfo{author}{\bibfnamefont{W.}~\bibnamefont{Willinger}},
  \bibnamefont{and} \bibinfo{author}{\bibfnamefont{D.~V.}
  \bibnamefont{Wilson}}, in \emph{\bibinfo{booktitle}{Proc. {ACM} {SIGCOMM}}},
  edited by \bibinfo{editor}{\bibfnamefont{D.~P.} \bibnamefont{Sidhu}}
  (\bibinfo{address}{San Francisco, California}, \bibinfo{year}{1993}), pp.
  \bibinfo{pages}{183--193}.

\bibitem[{\citenamefont{Willinger et~al.}(1997)\citenamefont{Willinger, Taqqu,
  Sherman, and Wilson}}]{willinger1997}
\bibinfo{author}{\bibfnamefont{W.}~\bibnamefont{Willinger}},
  \bibinfo{author}{\bibfnamefont{M.~S.} \bibnamefont{Taqqu}},
  \bibinfo{author}{\bibfnamefont{R.}~\bibnamefont{Sherman}}, \bibnamefont{and}
  \bibinfo{author}{\bibfnamefont{D.~V.} \bibnamefont{Wilson}},
  \bibinfo{journal}{{IEEE}\slash{ACM} Trans. on Networking}
  \textbf{\bibinfo{volume}{5}}, \bibinfo{pages}{71} (\bibinfo{year}{1997}).

\bibitem[{\citenamefont{Morris and Lin}(2000)}]{morris2000}
\bibinfo{author}{\bibfnamefont{R.}~\bibnamefont{Morris}} \bibnamefont{and}
  \bibinfo{author}{\bibfnamefont{D.}~\bibnamefont{Lin}}, in
  \emph{\bibinfo{booktitle}{Proc. {IEEE} {INFOCOM}}} (\bibinfo{year}{2000}),
  pp. \bibinfo{pages}{360--366}.

\bibitem[{\citenamefont{Ohira and Sawatari}(1998)}]{ohira1998}
\bibinfo{author}{\bibfnamefont{T.}~\bibnamefont{Ohira}} \bibnamefont{and}
  \bibinfo{author}{\bibfnamefont{R.}~\bibnamefont{Sawatari}},
  \bibinfo{journal}{Phys. Rev. E} \textbf{\bibinfo{volume}{58}},
  \bibinfo{pages}{193} (\bibinfo{year}{1998}).

\bibitem[{\citenamefont{Valverde and Sole}(2001)}]{valverde2001}
\bibinfo{author}{\bibfnamefont{S.}~\bibnamefont{Valverde}} \bibnamefont{and}
  \bibinfo{author}{\bibfnamefont{R.~V.} \bibnamefont{Sole}},
  \bibinfo{journal}{Physica A} \textbf{\bibinfo{volume}{289}},
  \bibinfo{pages}{595} (\bibinfo{year}{2001}).

\bibitem[{\citenamefont{Woolf et~al.}(2002)\citenamefont{Woolf, Arrowsmith,
  Mongrag\'{o}n, and Pitts}}]{woolf2002}
\bibinfo{author}{\bibfnamefont{M.}~\bibnamefont{Woolf}},
  \bibinfo{author}{\bibfnamefont{D.~K.} \bibnamefont{Arrowsmith}},
  \bibinfo{author}{\bibfnamefont{R.~J.} \bibnamefont{Mongrag\'{o}n}},
  \bibnamefont{and} \bibinfo{author}{\bibfnamefont{J.~M.} \bibnamefont{Pitts}},
  \bibinfo{journal}{Phys. Rev. E} \textbf{\bibinfo{volume}{66}},
  \bibinfo{pages}{046106} (\bibinfo{year}{2002}).

\bibitem[{\citenamefont{Erramilli et~al.}(1996)\citenamefont{Erramilli,
  Narayan, and Willinger}}]{erramilli1996}
\bibinfo{author}{\bibfnamefont{A.}~\bibnamefont{Erramilli}},
  \bibinfo{author}{\bibfnamefont{O.}~\bibnamefont{Narayan}}, \bibnamefont{and}
  \bibinfo{author}{\bibfnamefont{W.}~\bibnamefont{Willinger}},
  \bibinfo{journal}{{IEEE}\slash{ACM} Trans. on Networking}
  \textbf{\bibinfo{volume}{4}}, \bibinfo{pages}{209} (\bibinfo{year}{1996}).

\bibitem[{\citenamefont{Norros}(1994)}]{norros1994}
\bibinfo{author}{\bibfnamefont{I.}~\bibnamefont{Norros}},
  \bibinfo{journal}{Queueing Systems} \textbf{\bibinfo{volume}{16}},
  \bibinfo{pages}{387} (\bibinfo{year}{1994}).

\bibitem[{\citenamefont{Sahinoglu and Tekinay}(1999)}]{sahinoglu1999}
\bibinfo{author}{\bibfnamefont{Z.}~\bibnamefont{Sahinoglu}} \bibnamefont{and}
  \bibinfo{author}{\bibfnamefont{S.}~\bibnamefont{Tekinay}},
  \bibinfo{journal}{{IEEE} Communications Magazine}
  \textbf{\bibinfo{volume}{January}}, \bibinfo{pages}{48}
  (\bibinfo{year}{1999}).

\bibitem[{\citenamefont{Neidhardt and Wang}(1998)}]{neidhardt1998}
\bibinfo{author}{\bibfnamefont{A.~L.} \bibnamefont{Neidhardt}}
  \bibnamefont{and} \bibinfo{author}{\bibfnamefont{J.~L.} \bibnamefont{Wang}},
  in \emph{\bibinfo{booktitle}{Proceedings of the 1998 {ACM} {SIGMETRICS} joint
  international conference on Measurement and Modeling of Computer Systems}}
  (\bibinfo{year}{1998}), pp. \bibinfo{pages}{222--232}.

\bibitem[{\citenamefont{Mandelbrot}(1971)}]{mandelbrot1971}
\bibinfo{author}{\bibfnamefont{B.~B.} \bibnamefont{Mandelbrot}},
  \bibinfo{journal}{Water Resources Research} \textbf{\bibinfo{volume}{7}},
  \bibinfo{pages}{543} (\bibinfo{year}{1971}).

\bibitem[{\citenamefont{Davies and Harte}(1987)}]{davies1987}
\bibinfo{author}{\bibfnamefont{R.~B.} \bibnamefont{Davies}} \bibnamefont{and}
  \bibinfo{author}{\bibfnamefont{D.~S.} \bibnamefont{Harte}},
  \bibinfo{journal}{Biometrika} \textbf{\bibinfo{volume}{74}},
  \bibinfo{pages}{95} (\bibinfo{year}{1987}).

\bibitem[{\citenamefont{Paxson}(1997)}]{paxson1997}
\bibinfo{author}{\bibfnamefont{V.}~\bibnamefont{Paxson}},
  \bibinfo{journal}{Computer Comm. Rev.} \textbf{\bibinfo{volume}{27}},
  \bibinfo{pages}{5} (\bibinfo{year}{1997}).

\bibitem[{\citenamefont{Granger and Joyeux}(1980)}]{granger1980}
\bibinfo{author}{\bibfnamefont{C.~W.~J.} \bibnamefont{Granger}}
  \bibnamefont{and} \bibinfo{author}{\bibfnamefont{R.}~\bibnamefont{Joyeux}},
  \bibinfo{journal}{J. Time Ser. Anal.} \textbf{\bibinfo{volume}{1}},
  \bibinfo{pages}{15 } (\bibinfo{year}{1980}).

\bibitem[{\citenamefont{Wang}(1989)}]{wang1989}
\bibinfo{author}{\bibfnamefont{X.~J.} \bibnamefont{Wang}},
  \bibinfo{journal}{Phys. Rev. A} \textbf{\bibinfo{volume}{40}},
  \bibinfo{pages}{6647} (\bibinfo{year}{1989}).

\bibitem[{\citenamefont{Erramilli et~al.}(1995)\citenamefont{Erramilli, Singh,
  and Pruthi}}]{erramilli1995}
\bibinfo{author}{\bibfnamefont{A.}~\bibnamefont{Erramilli}},
  \bibinfo{author}{\bibfnamefont{R.~P.} \bibnamefont{Singh}}, \bibnamefont{and}
  \bibinfo{author}{\bibfnamefont{P.}~\bibnamefont{Pruthi}},
  \bibinfo{journal}{Queueing Systems} \textbf{\bibinfo{volume}{20}},
  \bibinfo{pages}{171} (\bibinfo{year}{1995}).

\bibitem[{\citenamefont{Suetani and Horita}(1999)}]{suetani1999}
\bibinfo{author}{\bibfnamefont{H.}~\bibnamefont{Suetani}} \bibnamefont{and}
  \bibinfo{author}{\bibfnamefont{T.}~\bibnamefont{Horita}},
  \bibinfo{journal}{Phys. Rev. E} \textbf{\bibinfo{volume}{60}},
  \bibinfo{pages}{422} (\bibinfo{year}{1999}).

\bibitem[{\citenamefont{Horita and Suetani}(2002)}]{suetani2002}
\bibinfo{author}{\bibfnamefont{T.}~\bibnamefont{Horita}} \bibnamefont{and}
  \bibinfo{author}{\bibfnamefont{H.}~\bibnamefont{Suetani}},
  \bibinfo{journal}{Phys. Rev. E} \textbf{\bibinfo{volume}{65}},
  \bibinfo{pages}{056217} (\bibinfo{year}{2002}).

\bibitem[{\citenamefont{Ashwin et~al.}(2004)\citenamefont{Ashwin, Rucklidge,
  and Sturman}}]{ashwin2004}
\bibinfo{author}{\bibfnamefont{P.}~\bibnamefont{Ashwin}},
  \bibinfo{author}{\bibfnamefont{A.~M.} \bibnamefont{Rucklidge}},
  \bibnamefont{and} \bibinfo{author}{\bibfnamefont{R.}~\bibnamefont{Sturman}},
  \bibinfo{journal}{Physica D} \textbf{\bibinfo{volume}{194}},
  \bibinfo{pages}{30} (\bibinfo{year}{2004}).

\bibitem[{\citenamefont{Giampieri and Isola}(2005)}]{giampieri2005}
\bibinfo{author}{\bibfnamefont{M.}~\bibnamefont{Giampieri}} \bibnamefont{and}
  \bibinfo{author}{\bibfnamefont{S.}~\bibnamefont{Isola}},
  \bibinfo{journal}{Disc. and Cont. Dyn. Sys.} \textbf{\bibinfo{volume}{12}},
  \bibinfo{pages}{115} (\bibinfo{year}{2005}).

\bibitem[{\citenamefont{Barenco and Arrowsmith}(2004)}]{barenco2004}
\bibinfo{author}{\bibfnamefont{M.}~\bibnamefont{Barenco}} \bibnamefont{and}
  \bibinfo{author}{\bibfnamefont{D.}~\bibnamefont{Arrowsmith}},
  \bibinfo{journal}{Dynamical Systems} \textbf{\bibinfo{volume}{19}},
  \bibinfo{pages}{61} (\bibinfo{year}{2004}).

\bibitem[{\citenamefont{Riedi}(2003)}]{riedi2003}
\bibinfo{author}{\bibfnamefont{R.~H.} \bibnamefont{Riedi}}, in
  \emph{\bibinfo{booktitle}{Theory And Applications Of Long-Range Dependence}},
  edited by \bibinfo{editor}{\bibfnamefont{P.}~\bibnamefont{Doukhan}},
  \bibinfo{editor}{\bibfnamefont{G.}~\bibnamefont{Oppenheim}},
  \bibnamefont{and} \bibinfo{editor}{\bibfnamefont{M.~S.} \bibnamefont{Taqqu}}
  (\bibinfo{publisher}{Birkh{\"a}user}, \bibinfo{year}{2003}), pp.
  \bibinfo{pages}{625--716}.

\bibitem[{\citenamefont{Riedi et~al.}(1999)\citenamefont{Riedi, Crouse,
  Ribeiro, and Baraniuk}}]{riedi1999}
\bibinfo{author}{\bibfnamefont{R.~H.} \bibnamefont{Riedi}},
  \bibinfo{author}{\bibfnamefont{M.~S.} \bibnamefont{Crouse}},
  \bibinfo{author}{\bibfnamefont{V.~J.} \bibnamefont{Ribeiro}},
  \bibnamefont{and} \bibinfo{author}{\bibfnamefont{R.~G.}
  \bibnamefont{Baraniuk}}, \bibinfo{journal}{{IEEE} Special Issue On
  Information Theory} \textbf{\bibinfo{volume}{45(April)}},
  \bibinfo{pages}{992} (\bibinfo{year}{1999}).

\bibitem[{\citenamefont{Taqqu}()}]{taqquwww}
\bibinfo{author}{\bibfnamefont{M.~S.} \bibnamefont{Taqqu}},
  \emph{\bibinfo{title}{Murad {S}. {T}aqqu's {H}omepage: \\ {\tt
  math.bu.edu/individual/murad/home.html}}}.

\bibitem[{\citenamefont{Taqqu et~al.}(1995)\citenamefont{Taqqu, Teverovsky, and
  Willinger}}]{taqqu1995}
\bibinfo{author}{\bibfnamefont{M.}~\bibnamefont{Taqqu}},
  \bibinfo{author}{\bibfnamefont{V.}~\bibnamefont{Teverovsky}},
  \bibnamefont{and}
  \bibinfo{author}{\bibfnamefont{W.}~\bibnamefont{Willinger}},
  \bibinfo{journal}{Fractals} \textbf{\bibinfo{volume}{3}},
  \bibinfo{pages}{785} (\bibinfo{year}{1995}).

\bibitem[{\citenamefont{Taqqu and Teverovsky}(1997)}]{taqqu1997}
\bibinfo{author}{\bibfnamefont{M.~S.} \bibnamefont{Taqqu}} \bibnamefont{and}
  \bibinfo{author}{\bibfnamefont{V.}~\bibnamefont{Teverovsky}},
  \bibinfo{journal}{Stochastic Models} \textbf{\bibinfo{volume}{13}},
  \bibinfo{pages}{723} (\bibinfo{year}{1997}).

\bibitem[{\citenamefont{Bardet et~al.}(2003)\citenamefont{Bardet, Lang,
  Oppenheim, Phillipe, Stoev, and Taqqu}}]{bardet2003}
\bibinfo{author}{\bibfnamefont{J.-M.} \bibnamefont{Bardet}},
  \bibinfo{author}{\bibfnamefont{G.}~\bibnamefont{Lang}},
  \bibinfo{author}{\bibfnamefont{G.}~\bibnamefont{Oppenheim}},
  \bibinfo{author}{\bibfnamefont{A.}~\bibnamefont{Phillipe}},
  \bibinfo{author}{\bibfnamefont{S.}~\bibnamefont{Stoev}}, \bibnamefont{and}
  \bibinfo{author}{\bibfnamefont{M.~S.} \bibnamefont{Taqqu}}, in
  \emph{\bibinfo{booktitle}{Theory and Applications of Long-Range Dependence}},
  edited by \bibinfo{editor}{\bibfnamefont{P.}~\bibnamefont{Doukhan}},
  \bibinfo{editor}{\bibfnamefont{G.}~\bibnamefont{Oppenheim}},
  \bibnamefont{and} \bibinfo{editor}{\bibfnamefont{M.~S.} \bibnamefont{Taqqu}}
  (\bibinfo{publisher}{Birkh{\"a}user}, \bibinfo{year}{2003}), pp.
  \bibinfo{pages}{557--577}.

\bibitem[{\citenamefont{Mandelbrot and Wallis}(1969)}]{mandelbrot1969}
\bibinfo{author}{\bibfnamefont{B.~B.} \bibnamefont{Mandelbrot}}
  \bibnamefont{and} \bibinfo{author}{\bibfnamefont{J.~R.}
  \bibnamefont{Wallis}}, \bibinfo{journal}{Water Resources Research}
  \textbf{\bibinfo{volume}{5}}, \bibinfo{pages}{228} (\bibinfo{year}{1969}).

\bibitem[{\citenamefont{Geweke and Porter-Hudak}(1983)}]{geweke1983}
\bibinfo{author}{\bibfnamefont{J.}~\bibnamefont{Geweke}} \bibnamefont{and}
  \bibinfo{author}{\bibfnamefont{S.}~\bibnamefont{Porter-Hudak}},
  \bibinfo{journal}{J. Time Ser. Anal.} \textbf{\bibinfo{volume}{4}},
  \bibinfo{pages}{221} (\bibinfo{year}{1983}).

\bibitem[{\citenamefont{Fox and Taqqu}(1986)}]{fox1986}
\bibinfo{author}{\bibfnamefont{R.}~\bibnamefont{Fox}} \bibnamefont{and}
  \bibinfo{author}{\bibfnamefont{M.~S.} \bibnamefont{Taqqu}},
  \bibinfo{journal}{The Annals of Statistics} \textbf{\bibinfo{volume}{14}},
  \bibinfo{pages}{517} (\bibinfo{year}{1986}).

\bibitem[{\citenamefont{Robinson}(1995)}]{robinson1995}
\bibinfo{author}{\bibfnamefont{P.~M.} \bibnamefont{Robinson}},
  \bibinfo{journal}{The Annals of Statistics} \textbf{\bibinfo{volume}{23}},
  \bibinfo{pages}{1630} (\bibinfo{year}{1995}).

\bibitem[{\citenamefont{Artuso and Cristadoro}(2003)}]{artuso2003}
\bibinfo{author}{\bibfnamefont{R.}~\bibnamefont{Artuso}} \bibnamefont{and}
  \bibinfo{author}{\bibfnamefont{G.}~\bibnamefont{Cristadoro}},
  \bibinfo{journal}{Physical Review Letters} \textbf{\bibinfo{volume}{90}},
  \bibinfo{pages}{244101} (\bibinfo{year}{2003}).

\bibitem[{\citenamefont{Feller}(1949)}]{feller1949}
\bibinfo{author}{\bibfnamefont{W.}~\bibnamefont{Feller}},
  \bibinfo{journal}{Trans. of the Amer. Math. Soc.}
  \textbf{\bibinfo{volume}{67}}, \bibinfo{pages}{94} (\bibinfo{year}{1949}).

\end{thebibliography}

\end{document}